# Hagen-Poiseuille Flow Linear Stability Paradox Resolving and Viscous Dissipative Mechanism of the Turbulence Emergence in the Boundary Layer

S.G. Chefranov [1)], A.G. Chefranov [2)]

[1)] A.M. Obukhov Institute of Atmospheric Physics RAS, Moscow, Russia;
e-mail: schefranov@mail.ru

[2)] Eastern Mediterranean University, Famagusta, North Cyprus; e-mail: Alexander.chefranov@emu.edu.tr

**Abstract.** In the linear theory of hydrodynamic stability up to now there exist examples of flows for which there is full quantitative distinction, as for cylindrical Hagen-Poiseuille (HP) flow in a pipe with round section, between theory conclusions and experimental data on the threshold Reynolds number $\mathrm{Re}_{th}$. In the present work, we show that to get a conclusion of linear instability of the HP flow for finite Reynolds numbers Re, it is necessary to abandon the use of traditional "normal" form of disturbances which assumes an opportunity of separation of variables describing disturbances variability depending on radial and longitudinal (along the pipe axis) coordinates. In the result of the absence of such variables separation, in the suggested linear theory, it is proposed to use Bubnov-Galerkin's approximation method modification that gives an opportunity to account longitudinal variability periods distinctions for different radial modes defined a priori in the result of standard Galerkin-Cantorovich's method to the equation of evolution of extremely small axially symmetric velocity field tangential component disturbances. We found that when considering even two linearly interacting radial modes for the HP flow, linear instability is possible only when there exists mentioned above conditionally periodic longitudinal along the pipe axis disturbance variability when $\mathrm{Re}_{th}(p)$ very sensitively depends on the ratio $p$ of two longitudinal periods each of which describes longitudinal variability for its own radial mode only. Obtained for the HP flow linear instability realization minimal value $\mathrm{Re}_{th}(p) \approx 448$ (when $p \approx 1.527$) quantitatively agrees with the Tolmin-Shlihting waves in the boundary layer mergence, where also $\mathrm{Re}_{th} = 420$. We get good quantitative agreeing of the phase velocity values of the considered vortex disturbances with experimental data on the fore and rear fronts of the turbulent "puffs" spreading along the pipe axis.

PACS: 47.20.Ft, 47.27.Cn, 47.27.nf

# Hagen-Poiseuille Flow Linear Stability Paradox Resolving and Viscous Dissipative Mechanism of the Turbulence Emergence in the Boundary Layer

S.G. Chefranov, A.G. Chefranov

**Introduction.** Fundamental and applied problem of defining of the turbulence emergence mechanism for the Hagen-Poiseuille (HP)[1] flow more than century is left mysterious because of the linear stability paradox of the flow with respect to extremely small by amplitude disturbances for any Reynolds number value $\mathrm{Re} = \frac{V_{\max} R}{\nu}$ (where $V_{\max}, \nu, R$ are the maximal HP flow near axis velocity, kinematic viscosity coefficient, and pipe radius respectively) [1-4]. Obvious contradiction with experiments corresponding to the paradox now is used to be coped with based on an assumption of permissibility of the HP flow instability with respect to disturbances having sufficiently large finite amplitude strict non-linear mechanism only [5-10]. The basis for such the assumption (see [3, 4]) gives up to now only by one side interpretation of experiments [11] in which many fold increase of the threshold Reynolds number value to $\mathrm{Re}_{th} = 100000$ is achieved due to the increase of the level of smoothness of the streamlined pipe surface. In this interpretation, only correlation between the surface smoothness increase and resultant decrease of the average amplitude of the original disturbances is taken into account. At the same time, noted even by O.Reynolds [1] extremely high sensitivity of the value of $\mathrm{Re}_{th}$ to the initial disturbance does not exclude possibility of impact on $\mathrm{Re}_{th}$ of not only amplitude but also space-time characteristics of the disturbances also caused by non-ideal smoothness of the streamlined surface. Actually, for example in the experiment [12], it is found that under the fixed amplitude of artificially excited disturbances, instability of the HP flow emerges only in some definite narrow range of the disturbances' frequencies.

In the present work, we show that possibility of linear absolute (i.e., non-convective [4]) instability of the HP flow is defined by the value of complementary to the Reynolds number *Re* control parameter $p$, which characterizes frequency-wave number features of the disturbances and affects on the value of the threshold Reynolds number $\mathrm{Re}_{th}(p)$ independently from the amplitude of the initial disturbances. Role of such complementary parameter for simple example of the HP flow modification (when the presence of a cylinder rod on the pipe axis leads to the appearance of linear instability region for finite Re, see [13]) can play the ratio of the radii of the external and near-axis cylinders. Similarly, complementary parameters are easily introduced also in other HP flow modifications – in the cases of the flow in the pipe with elliptic cross section [14], flow in the rotating pipe [15, 16], and even for a flow transferring particles of finite size in a pipe [17]. In all this examples, there already exists complementary to the Reynolds number control parameter $p$ and the linear stability theory paradox is absent. The mention examples explicitly specify that "avoidance or circumvention" (see [5]) of the HP flow linear stability paradox due to the consideration of strict finite amplitude only mechanism of instability of the flow "hardly can satisfy anybody" [18].

Introduction of such a complementary parameter $p$ for the HP flow is already not as obvious as for the HP flow modifications in [13 – 17]. It however is performed below on the base of pointed by O. Reynolds [1] (and then by W. Heisenberg also for the flat Poiseuille flow,

[1] HP flow is by definition a laminar stationary flow of the uniform viscous fluid along the static straight linear and unbounded in length pipe with the round, and the same along the whole pipe axis, cross section

see in [4, 6]) concept on dissipative instability mechanism[2] of the HP flow related with the action of molecular viscosity $\nu$ near the very solid boundary. According to [1], the mechanism manifests itself in the form of spontaneous one-step emergence for $\mathrm{Re} > \mathrm{Re}_{th}$ of vortexes having character size $L_\nu$, «..that is already not growing as it was expected with the growth of the velocity amplitude [1]». That is why, the value $L_\nu$ must significantly differ also from the length scale $l_\nu = \nu / V_{\max}$ that leads to the Reynolds number defining as $\mathrm{Re} = R / l_\nu$ and explicitly depending on the stream velocity maximal amplitude value. Such scale $L_\nu$ seemingly related also with the level of the streamlined pipe surface smoothness, may together with the of radius *R* define not amplitude only but also frequency-wave number initial disturbance parameters, for example, their longitudinal along the pipe axis (axis *z*) spatial periods. The ratio of the periods $p = \frac{L_\nu}{R}$ as it is shown below is a new complementary parameter defining the HP flow linear instability threshold with respect to extremely small by amplitude vortex disturbances. Note that for any Reynolds number value, *p* can vary in vast ranges from $p<<1$ to $p>>1$.

It is suggested here to use disturbance structure representation in the form of two radial modes each of which has its own period of longitudinal variability differing from that of the other mode. Such representation corresponds to the observed conditionally periodic Tolmin - Shlihting (TS) waves emergence of which (caused also by near boundary action of the molecular viscosity) precedes blow-like emergence of the turbulence in the near-boundary layer. [23-25]. Besides that, even in [2, 26], it is noted that usually considered in the linear stability theory "normal" periodic by $z$ disturbances fields obviously don't correspond to the structures observed in the experiments, for which different longitudinal periods for different radial modes are characteristic..

In the present work, it is shown that leaving off the assumption on separation of the longitudinal and radial variables defining spatial disturbance variability leads now to the finite value of the minimal threshold Reynolds number $\mathrm{Re}_{th} \approx 448$ (for $p \approx 1.53..$). Close to it threshold Reynolds number value is characteristic also for the observed threshold for the transition from the laminar resistance law to another type one [2, 27] and for the conditions of excitation of TS waves in a boundary layer [25]. We have conducted comparison of the considered theory with the experimental data for the flow in the pipe [28-30] and also with the conclusions of the stability theory (Tolmin - Shlihting and Lin C. C.) and experimental data on the stability of laminar near boundary layer [31]. We obtained correspondence not only of the quantitative values of the critical Reynolds number for linear exponential instability for the HP flow and for TS waves excitation (where also $\mathrm{Re}_{th} = 420$), but also similar shapes of instability regions (bounded by the curves of neutral stability). This also confirms expected above similarity of their viscous dissipative realization mechanisms.

## Linear instability of the HP flow

1. Let us consider known (see [4]) representation of the HP flow in the cylindrical reference frame $(z,r,\varphi)$: $V_{0r} = V_{0\varphi} = 0, V_{0z} = V_{\max}(1 - \frac{r^2}{R^2})$, where $V_{\max} = \frac{R^2}{4\rho\nu}\frac{\partial p_0}{\partial z}$, the fluid density $\rho = const$, $\frac{\partial p_0}{\partial z}$ is the constant value of the pressure gradient $p_0$ along the axis of the pipe of radius $R$, and $\nu$ is the coefficient of kinematic fluid viscosity.

[2] Such a mechanism is natural realized in the systems having disturbances with negative energy [19-22], for example, for threshold emergence of vortexes (rotons) in the flow of super-fluid helium in a capillary [19].

In the axially symmetric case (i.e. for extremely small disturbances not depending on $\varphi$) linear instability of the HP flow can be defined by the tangent velocity component $V_\varphi$ only, which before entering the nonlinear stage of the evolution is not related with the other components of the disturbed velocity field and pressure field, and meets the following equation:

$$\frac{\partial V_\varphi}{\partial t} + V_{0z}(r)\frac{\partial V_\varphi}{\partial z} = \nu(\Delta V_\varphi - \frac{V_\varphi}{r^2}) \ , \qquad (1)$$

where $\Delta$ is the three-dimensional Laplace operator.
Only when $\nu \neq 0$ in (1), it is possible to expect realization of the HP flow viscous dissipative instability for some above threshold Reynolds numbers $Re > \mathrm{Re}_{th}$, where $Re = \frac{V_{\max} R}{\nu}$.
Actually, when $\nu = 0$, the right-hand side of (1) also turns to zero. In that case, it takes place only convective disturbance transfer without change of their form and amplitude with time, and application of any averaging procedure over the longitudinal variable can't qualitatively change the conclusion. Let's note that due to the consideration in (1) of only velocity field tangential component disturbances, it is automatically provided identical conservation of the mass stream through the cross section of the pipe for the superposition of the main flow and disturbance field. Natural emergence of such disturbances in an axially symmetrical stream really may be hindered although it can't be fully excluded due to the possibility of presence of corresponding randomly-non-uniform roughness of the streamlined pipe surface. In the laboratory simulation of the HP flow such disturbances can be artificially created (see [12]). Let's note also that combination of the main stream and considered disturbances field has non zero value of integral helicity.
Let's seek solution of (1) using Galerkin-Kantorovich method and describing $V_\varphi$ as follows

$$V_\varphi = V_{\max}\sum_{n=1}^{N} A_n(z,t) J_1(j_{1,n}\frac{r}{R}) \ , \qquad (2)$$

where $V_\varphi$ in (2) for any $A_n$ meets necessary boundary conditions on $r$ ($V_\varphi < \infty$ for $r = 0$ and $V_\varphi = 0$ for $r = R$), since $J_1$ is the Bessel function of the first order, and $\gamma_{1,n}$ are its zeroes, i.e. $J_1(\gamma_{1,n}) = 0$ for any integer n. In (2) velocity field on the pipe axis not only satisfies the boundary condition of restricted $V_\varphi < \infty$, but also turns to zero. Correspondingly, in (2), instead of the Bessel functions of the first order there might be used also Bessel functions of other orders for which it is provided only zero boundary condition on the pipe boundary, and on the pipe axis, velocity field disturbances would be non zero valued. For the disturbance representation in the form (2), only vortex field radial component $\omega_r = -\frac{\partial V_\varphi}{\partial z}$ shall turn to zero, and the value of the longitudinal vortex field component $\omega_z = \frac{1}{r}\frac{\partial r V_\varphi}{\partial r}$ on the pipe boundary has already non zero value that corresponds to the character of forming of the vortex disturbances due to interaction of the stream with the solid pipe wall caused by viscosity forces.

For the coefficients $A_n$, characterizing amplitudes of the linearly interacting disturbance field radial modes, from (1), (2), we get the following system of equations in dimensionless form:

$$\frac{\partial A_m}{\partial \tau} + j_{1,m}^2 A_m - \frac{\partial^2 A_m}{\partial x^2} + \mathrm{Re}\sum_{n=1}^{N} P_{nm}\frac{\partial A_n}{\partial x} = 0, \qquad (3)$$

where $m = 1,2,..,N, \tau = \frac{t\nu}{R^2}, x = \frac{z}{R}$. In (3), constant coefficients $P_{nm}$ have the form

$$P_{nm} = \frac{2}{J_2^2(j_{1,m})} \int_0^1 dy\, y(1-y^2) J_1(j_{1,n}y) J_1(j_{1,m}y) \ , \qquad (4)$$

where $J_2$ are the Bessel functions of the second order and the linear with respect to $y$ term under the integral sign yields in $P_{nm}$ the contribution in the form of unity matrix $\delta_{nm} = \begin{cases} 1, n = m \\ 0, n \neq m \end{cases}$. For $N = 1$ in (3), the last term can be excluded by Galileo transformation and hence for $N = 1$, there is no opportunity of the global absolute instability of the HP flow. In that relation, we shall consider (3) in the simplest non-trivial case $N = 2$, that allows already to resolve the HP flow linear stability paradox and leads to the conclusions quantitatively agreeing with the experimental data [29, 31].

2. As it was already noted, observed in the experiment field structures do not correspond to strictly periodic along the pipe axis disturbances changes (see above and [2,26]). More over, in [26], it is noted that different radial modes (defining dependence of the disturbances on the radial coordinate) have corresponding differing each from the other variability periods along the pipe axis. This behavior of the observed disturbances change can be modeled with the help of the use in the representation of the system (3) solution an assumption on the difference of the longitudinal periods along the pipe axis for radial modes with different values of index m. Such a requirement corresponds to the introduction for each of these modes of its own, independent from the other modes, periodical boundary condition on x. In the result, there emerges necessity in the use of the adequate to the pointed boundary conditions Galerkin's approximation of the system (3) solution obtain. Let in (3), for $N = 2$, amplitudes $A_1$ and $A_2$ have the form of the running waves with different periods along the pipe axis:

$$A_1 = A_{10} e^{\lambda\tau + ix2\pi\alpha}, \ A_2 = A_{20} e^{\lambda\tau + ix2\pi\beta}, \qquad (5)$$

where $A_{10}$ and $A_{20}$ are the constant values. Meanwhile, complementary to the Reynolds number $\mathrm{Re}$ control parameter can be defined as $p = \frac{\alpha}{\beta}$ for any $\alpha$ and $\beta$.

For N=2, system (3) after substitution in it of the solution representation (5) has the form of a system of two discrepancies:

$$(\lambda + \gamma_{1,1}^2 + 4\pi^2\alpha^2 + i2\pi\alpha P_{11}\,\mathrm{Re}) A_{10} e^{ix2\pi\alpha} + iA_{20} e^{ix2\pi\beta} 2\pi\beta P_{21}\,\mathrm{Re} = 0 \qquad (3.1)$$

$$iA_{10} e^{ix2\pi\alpha} 2\pi\alpha P_{12}\,\mathrm{Re} + (\lambda + \gamma_{1,2}^2 + 4\pi^2\beta^2 + i2\pi\beta P_{22}\,\mathrm{Re}) A_{20} e^{ix2\pi\beta} = 0 \qquad (3.2)$$

When considering the system (3.1), (3.2), we use the Bubnov-Galerkin's method in the modification applied here to the very system of two discrepancies that allows taking into account the supposed absence of the spatial variables separation when there exists corresponding to the observations distinctions of the longitudinal variability periods for different radial modes. In the result, the system (3.1), (3.2) is transformed into a uniform system with constant coefficients for the unknown values $A_{10}$ and $A_{20}$. From the condition of solvability of the system for the non zero $A_{10}$ and $A_{20}$, we define the value of the exponent $\lambda = \lambda_1 + i\lambda_2$ depending on dimensionless parameters $\mathrm{Re}$, $p$ and $\beta$ (see (A.1) and (A.2) in Appendix).

Used in the present work modification of weighted discrepancies according to Bubnov-Galerkin method when considering discrepancies (3.1), (3.2) consists in the following. Let's multiply discrepancy (3.1) by $\exp(-ix2\pi\alpha)$ and take integral $\alpha\int_0^{1/\alpha}dx$ for each its term. Similarly, for discrepancy (3.2), let's multiply it by $\exp(-ix2\pi\beta)$ and average it taking integral $\beta\int_0^{1/\beta}dx$.

Such an application of the Galerkin's approximation on the longitudinal variable takes into account absence of any domination of one of the longitudinal periods characterizing the solution representation in the form (5). If (3.1) and (3.2) to average over one ant the same period of the longitudinal variability (with multiplication of (3.1) and (3.2) by one and the same multiplier), then in that case the HP flow linear instability will be absent. And there will not be distinction from the case when the spatial variables separation takes place and the both radial modes have one and the same longitudinal variability period. It means that for such averaging, actually, there will be neglected the difference of the radial modes longitudinal variability periods suggested in the solution representation (5). The pointed above procedure of the use of the Galerkin's approximation with averaging of the first and second discrepancies of the system (3.1) and (3.2) over different periods looks the most natural but does not exclude the possibility of application of other averaging procedures also. For example, it is possible to average (3.1) and (3.2) over one and the same period $\gamma = 1/\sqrt{\alpha\beta}$, but in that case, it is necessary to multiply (3.1) by $\exp(ix2\pi\gamma)$, and (3.2) by $\exp(-ix2\pi\gamma)$ (or vice versa). Obtained thus threshold (for the HP flow linear instability) Reynolds number is found out insensitive to the pointed variation in the averaging procedure (when corresponding value $\mathrm{Re}_{th}$ does not depend on which of the discrepancies, (3.1) or (3.2), is multiplied by $\exp(-ix2\pi\gamma)$, and which by $\exp(ix2\pi\gamma)$ ). More detailed consideration of the possible quantitative conclusions about the HP flow instability dependency on the details of the applied averaging procedure may be conducted elsewhere. Meanwhile, it is interesting also to consider more general cases with N>2, as well as the use instead of (5) representations for radial modes longitudinal variability via conditionally periodic functions now for each of the radial modes (not coinciding each with the other for different radial modes). Here we note only permissibility of the very fact of existence of linear exponential (not algebraic, with the power law of growth with time) instability of the HP flow stated in the present work. Let's note that even if in the system (3), to turn the kinematic viscosity coefficient to zero, then this does not exclude as for (1) possibility of the HP flow linear instability realization. It is related with the fact that already the very inference of (3) from (1) for N>1 is based on the finiteness of the kinematic viscosity coefficient in (1).

3. The condition of existence of linear exponential instability with $\lambda_1 > 0$ in (5) has the form (A.3). For $\mathrm{Re} >> 1$, (A.3) may be reduced to (A.4). Meanwhile, the value $\beta$, defined in (A.5), minimizes the expression for $\mathrm{Re}_{th}$ in (A.4) and defines the following condition of the HP flow linear instability condition (giving the estimate from below of the exact value $\mathrm{Re}_{th}$, defined from (A.3)):

$$\mathrm{Re} > \mathrm{Re}_{th} = \frac{\pi^2(1-p)^2 F^{1/2}}{P_{12}P_{21}p^2|S|}, \tag{6}$$

where $S = \sin\pi p \sin\frac{\pi}{p}\sin\pi(\frac{1}{p}+p)$, $B = \frac{S}{|S|}(pP_{11} - P_{22})$,

$F = (\gamma_{1,2}^2 + \gamma_{1,1}^2)(1+p^2)A^2 + (\gamma_{1,2}^2 - \gamma_{1,1}^2)(1-p^2)B^2 + 2AB(\gamma_{1,2}^2 - p^2\gamma_{1,1}^2)$,

$$A^2 = B^2 - \frac{4S\,P_{12}P_{21}p^2 ctg\pi(p+\frac{1}{p})}{\pi^2(1-p)^2}$$ for $P_{11}, P_{22,} P_{12}, P_{21}$ from (4) , any $p$ and under condition that here $A^2 > 0$.

For negative values , $A^2 < 0$, it is already necessary instead of (A.4) and (6) to use the general condition (A.3), under which $\lambda_1 > 0$ in (5) and it takes place exponential growth with time of the velocity field tangential component disturbance amplitude.

In the condition (6), providing realization of the HP flow linear instability realization Reynolds number threshold value can ten to infinity $\mathrm{Re}_{th} \to \infty$ only for such $p$, for which the denominator in (6) turns to zero. It takes place for $S = 0$, when the value of the ratio of longitudinal periods is equal to one of the following irrational numbers $p = p_k = k, p = p_{1/k} = \frac{1}{k}$, or equals to one of the irrational numbers defined by the following equality $p = p_{\sqrt{k}} = \frac{k+1 \pm \sqrt{(k+1)^2 - 4}}{2}$ for any integer k ($k = 1,2,..$). For $p$, related to the intervals of variability $p$ between any two neighboring values $p_k, p_{1/k}, p_{\sqrt{k}}$, the value $\mathrm{Re}_{th}$ in (6) is a function of $p$, having one local minimum on each of the mentioned intervals (see Fig. 1,b). And the value of the absolute minimum $\tilde{\mathrm{Re}}_{th}^{\min} \approx 442$ in (6) is reached for $p \approx 1.53..$, close to the value of the "golden" ratio $p_g = \frac{1+\sqrt{5}}{2} \approx 1.618..$ (i.e., the limit of the infinite sequence of the ratios of two neighboring Fibonacci numbers each of which is equal to the sum of two previous numbers: 1, 2, 3, 5, 8, 13, 21, etc.). For the same $p$, from the exact condition (A.3), we get close value of the absolute minimum $\tilde{\mathrm{Re}}_{th}^{\min} \approx 448$ (see also Table in the Appendix where conclusions on the base of (A.3) and (6) are compared).

**Comparison with experimental data and results of TWs (travelling waves) numerical simulation**

1. The found value $\tilde{\mathrm{Re}}_{th}^{\min} \approx 448$corresponds to the interval of values $\mathrm{Re} \in 300 \div 500$, noted in experimental observation of the threshold transition of the laminar resistance law (for a flow in the pipe) to another already non laminar (but yet not obviously turbulent) resistance mode [2, 27] and for Tolmin-Shlihting (TS) waves in the near wall region of the boundary layer [25]. Observed in [1] and other experiments (see references in [29, 30]) unusual sensitivity of the value $\mathrm{Re}_{th}$ to the initial disturbances, actually, corresponds to the obtained in (6) dependency of $\mathrm{Re}_{th}$ on $p$, when, for example, $\mathrm{Re}_{th}$ in (6) changes nearly 600 times only when $p$ changes from the value 0,12 to the value 0,11. Neighboring local minima of $\mathrm{Re}_{th}$ in (6) also may significantly differ each from the other. For example, for the value $p \approx 2.23$, we have in (6) $\mathrm{Re}_{th} \approx 1982$, and for the value $p \approx 3.86$, already we get $\mathrm{Re}_{th} \approx 84634$. In the scaled form, fragments of the neutral curve, corresponding to the condition (6) (see Fig. 1b), are given on Fig. 1a) in the form of dependency of the value *1/2p* on *Re*. They are plotted on the taken from the paper [31] figure (see Fig.12 in [31]), on which theoretical (of Lin and Shlihting) neutral curves and respective experimental data, related with the defining of instability emergence threshold in a boundary layer, are given. Obvious correspondence of the results following from Fig. 1a) allows to make the conclusion also about similarity of the linear dissipative instability mechanisms realized for the HP flow (when meeting the condition (6) or (A.3)), as well as for Tolmin-Shlihting waves excitation in a boundary layer.

2. Conditionally periodic with respect to $z$ structure of the initial disturbances field $V_{\varphi}$ in the representation of the solution (1) in the form (2) - (5) agrees with the observed ( in [29] ) wave velocity field tangential component disturbances changes along the pipe axis. It is especially obvious near very turbulence dying threshold for $\mathrm{Re} \approx 1750$, when on Fig. 5d in [29], it is possible to recognize pairs of characteristic longitudinal variability periods ratios of which are close to the values $p \approx \frac{8}{5} = 1.6$ and $p \approx \frac{13}{8} = 1.625$ are close to the value of the "golden" value of the periods ratio $p_g = 1.618...$ .

3. On the Fig. 2a) (where Fig. 4 from [30] is used as a basic), dependency on $\mathrm{Re}$ of the turbulent spot rear front constant velocity observed for the flow in a pipe is given as well as respective experimental data from [28], scaled by the average flow velocity. There, data are given from [28], corresponding to the turbulent spot rear front velocity description (blue triangles), as well as that of the fore front (light triangles). Also, there, results are given following from the TW in the pipe non-linear theory [8], and also conclusions of the present work for the phase velocity $V_{\beta} = -\frac{\lambda_2 V_{\max}}{2\pi\beta \mathrm{Re}}$ (scaled by the HP flow average velocity: $V_{av} = \frac{V_{\max}}{2}$). The value of the phase velocity is defined in (A.6) from $\lambda_2$ in (A.2) for the neutral curve (i.e. under condition $\lambda_1 = 0$). In the present theory, the estimates V=1.4 and V=0.8 for the velocities of the fore (для скоростей переднего (leaving the average stream behind) and rear (remaining behind the average stream) fronts of the vortex disturbance in the units of the HP flow average velocity. Experimental data [28, 30] give respective values *V = 1.2* and *V= 0.75*, and numerical calculations on the base of the non-linear theory [8] yield the possibility of the change of *V* from *1.55* to *0.95*. In the result, conclusions of the present linear theory lead to the better agreeing with the experimental data compared with those of the non-linear theory [8], especially in the estimate of the velocity of remaining of the velocity of spreading of the vortex disturbances flow the average velocity of the main flow which for the present work is $0.2V_{av}$, for the experiment is $0.25V_{av}$, and for the non-linear theory is $0.05V_{av}$.

Thus, conducted comparison of the suggested theory with the observation data and results of the non-linear theory shows that these data and the conclusions of the present HP flow linear instability theory are in satisfactory quantitative and qualitative agreement each with the other.

**Discussion of the stated HP flow dissipative linear instability mechanism and conclusions**

Obtained in the present work new conclusion on possibility of the proof of existence of linear instability for the HP flow is based on the analysis of the system (3) inferred from the initial equation (1) for the evolution of the velocity field tangential component disturbances under condition that the right hand side of (1) is non-zero due to the finiteness of the kinematic viscosity coefficient. Otherwise, when the coefficient is equal to zero, from (1), in principle, it is not possible to get evolution equations for linearly interacting each with the other radial modes and come to the conclusion of the HP flow linear instability. That is why, the mechanism of the found out HP flow linear instability can be called dissipative, and the very instability to consider as the dissipative instability.

Earlier, such hydrodynamic instability dissipative mechanism was considered in the works of L. Prandtl (1921-22) when investigating laminar boundary layer stability, and also of W. Heisenberg (1924) and Lin C. C. (1944-45) when establishing flat Poiseuille flow linear instability (see also [23] and references therein). Qualitative explanation of the physical sense and possibility of appearance off dissipative instability in the problems of hydrodynamic stability are discussed in [18] on the base elementary accounting of the delay effects on example off an oscillator with the friction linear with respect to velocity. Substantial understanding of the phenomenon of the dissipative instability for HP and other flows near solid boundary surface may also be obtained using a method similar to the one suggested by L.D. Landau in [19] for estimation of the critical velocity of motion of the super-fluid liquid in a capillary. In [19], from the condition of negativity of the energy of an elementary vortex disturbance when for the velocities exceeding that critical value due to the viscous interaction of the stream with the capillary wall, there emerges a vortex disturbance (roton) destroying the laminar super-fluid state of the liquid motion. For the HP flow, for example, also it is interesting to conduct similar to [19 – 22] research aiming defining conditions for realization of the dissipative instability related with the threshold character of the emerging vortex disturbance energy becoming negative valued (in an appropriate inertial reference frame) when exceeding some definite critical Reynolds number.

Let us note, however, that in the present work, to get for the HP flow the conclusion on linear instability, accounting of finiteness of the viscosity is important only on the stage of getting, from the equation (1), the system (3) that defines evolution with time and along the pipe axis, for $N>1$, of the linearly interacting radial disturbance modes. Already in (3) or in (3.1), (3.2), it is possible to consider the limit of infinitely large values of *Re*, corresponding to the ideal liquid with zero viscosity. Meanwhile, it is important only to preserve the suggested above consideration of the linear hydrodynamic stability problem for the very case of the boundary conditions individually defined for each of the both considered (for $N=2$) radial disturbance modes. Respective averaging procedure in (3.1), (3.2), as it was noted above, shall not neglect noted distinction of the longitudinal variability periods for the radial modes under consideration. Only in that case, it is preserved the obtained conclusion about possibility of the HP flow exponential instability but now instead of two control parameters, (*Re, p*), defining the instability region (depending on the wave number $\beta$), in the limit of zero viscosity, there will be left only the parameter *p*. In the present work, such a limit of zero viscosity for the system (3) (and (3.1), (3.2), respectively) was not considered. Such an investigation on the base of the linear system of interacting radial modes (3) for $N=2$ may be interesting in relation with available works on simulation of the processes of instability formation in the flow in a pipe based on the use of the concept of ideal (non-viscous) disturbances describing non-linear pair-wise interacting TWs with small but finite amplitude [32-34]. At the same time, we show in the present work that for the finite value of the kinematic viscosity coefficient, consideration of the limit $Re>>1$, yielding the formula (6) for estimation of the minimal threshold Reynolds number gives not large difference in the value of the estimate (since it was obtained the value $\mathrm{Re}_{th}=442$) from the exact formula (A.3), where $\mathrm{Re}_{th}=448$.

Let us note that the considered double vortex-wave structure of the spatial disturbance field variability is in the qualitative agreement with the data of laboratory [35] and numeric [36, 37] experiments in the pipe. This is witnessed also by the conducted in the previous paragraph comparison of the conclusions of the present theory with the experimental data and results of numerical modeling of instability development for the flow in the pipe. And in [36], for example, there were obtained estimates of the turbulent spot phase velocity $V=0.9$ and $V=1.1$ (in the units of the flow in the pipe average velocity), similar to the presented above.

To obtain the conclusion about possibility of the HP flow linear instability as it was noted above, it is important to apply to the discrepancies system (3.1), (3.2) such an averaging procedure that corresponds to the supposed distinction in the periodic boundary conditions of the different radial modes. The radial modes have differing each from the other longitudinal variability periods that corresponds to the use of representation (5) for them. And according to (6), linear exponential instability is found to be possible not only for almost all irrational values of ratios of such longitudinal periods *p*, but also for the rational values of $p$ , not coinciding with $p = p_k = k$ or $p_{1/k} = 1/k$ for integer *k ( k=1, 2, 3..)*. An exemption from all possible irrational values *p* constitute only defined from (6) irrational numbers $p = p_{\sqrt{k}} = \dfrac{k+1 \pm \sqrt{(k+1)^2 - 4}}{2}$, $k = 1,2,..$, for which, vice versa, $\mathrm{Re}_{th} \to \infty$ in (6) and the HP flow linear instability can't be realized.

Let us give an example showing that not for any procedure of averaging of discrepancies (3.1), (3.2), it is possible to get the HP flow linear instability condition. So, for example, for the rational value *p=3/2,* from (A.3) and (6), it follows the [possibility of the HP flow linear instability realization. That conclusion however can't be obtained if for (3.1) and (3.2) to use not the described above averaging procedure but to conduct averaging of the both discrepancies over the period $2/\beta$, that is the common period of the longitudinal variability for the both considered radial modes for *p=3/2*. For that, it is necessary to integrate $\beta/2 \int_0^{2/\beta} dx$ for each term in the left-hand side of (3.1) and (3.2) without multiplying them by any multiplier. For such averaging, the left hand sides of (3.1) и (3.2) now identically turn to zero for any value of the time evolution exponent coefficient $\lambda$. If the pointed averaging to conduct in (3) (for *N*=2) before substituting in (3) of the solution representation (5) (for *p=3/2*), then the both terms containing derivatives on *x* turn to zero. And the both disturbance amplitude values averaged over the period $2/\beta$, can only exponentially decay with time. Actually, for the pointed averaging procedure, it is explicitly neglected supposed distinction of the longitudinal variability periods for the considered two radial modes and it can't be defined the HP flow linear instability threshold. From the other side, the considered example (also for any other rational value *p*) means that in spite of the obtained in (6) and in (П.3) linear instability condition the vorticity disturbances spatial distribution may have itself such a character when in the different regions of the along the pipe axis there are regions with the different signs of the vortex field when the integral over the both regions yields zero integral value of the vortex disturbance field. That zero value of the integral vorticity is preserved also in the process of exponential growth in each of the sub-regions with the same vortex field sign (when realizing the HP flow linear instability for above critical Reynolds numbers defined in (6) and (A. 3)). At the same time, for any irrational value *p*, integral vorticity over any interval along the pipe axis already always is not identically zero that may provide also non zero value of the integral helicity for the combination of the main stream and disturbance field. For *p* not equal to the noted above values $p_{\sqrt{k}}$ , such an integral helicity can exponentially grow with time when realizing HP flow linear instability for $\mathrm{Re} > \mathrm{Re}_{th}$, where the threshold Reynolds number value $\mathrm{Re}_{th}$ is defined in (6) and (A.3).

The obtained conclusions of the present work allow filling the well-known gap in the non-linear theory [7, 8], when instead of the linear exponential instability up to now it was necessary to consider the stage of the seed algebraic instability (where small initial disturbances can only locally in time grow tending to zero for $t \to \infty$).

Let us note also that it is reasonable to revise also the mentioned above problems on linear stability for the flat Couette flow and the flat Poisuille flow on the base of accounting of the obtained in the present work conclusion about possibility of the HP flow linear exponential instability due to the consideration of differing from the "normal" longitudinal quasi-periodic disturbances. Meanwhile, it is important not the longitudinal disturbances quasi-periodicity by itself but formation of it due to the longitudinal periods distinctions for different basic (in that case, radial) modes formed only for the non zero fluid viscosity.

We are grateful to S.I. Anisimov, G.S. Golitsin. E.A. Novikov, and N.A. Inogamov for useful comments and interest to the work.

## *Appendix*

1. From (3) and (5) for $N = 2$ in Galerkin's approximation (representing dependence of disturbance on the longitudinal coordinate $z$, when in the general case $\alpha \neq \beta$ in (5)) we get for $\lambda = \lambda_1 + i\lambda_2$:

$$\lambda_1 = -\gamma_{1,1}^2 - 4\pi^2\beta^2 p^2 - \frac{1}{2}\left(a_1 \pm \frac{1}{\sqrt{2}} D_1^{1/2}\right) , \qquad \text{(A.1)}$$

$$\lambda_2 = -2\pi\beta p P_{11}\mathrm{Re} - \frac{1}{2}\left(a_2 \pm \frac{1}{\sqrt{2}} D_2^{1/2}\right) , \qquad \text{(A.2)}$$

where $D_1 = d_0^{1/2} + l$ , $D_2 = d_0^{1/2} - l$ , $l = a_1^2 - a_2^2 + 4c_1 \mathrm{Re}_1^2$

$a_1 = \gamma_{1,2}^2 - \gamma_{1,1}^2 + 4\pi^2\beta^2(1 - p^2)$, $a_2 = 2\pi\beta \mathrm{Re}(P_{22} - pP_{11})$,

$\mathrm{Re}_1^2 = \dfrac{\mathrm{Re}^2 P_{21}P_{12}p^2\beta^2}{(1-p)^2}$ , $d_0 = l^2 + 4(a_1 a_2 - 2\mathrm{Re}_1^2 d_1)^2$, $d_1 = -4S$ ,

$c_1 = -4S ctg\pi(p + \dfrac{1}{p})$, and $S$ is defined in (6).

The condition $\lambda_1 > 0$ leads to the inequality

$$(a\mathrm{Re} + b)^2 > c + \frac{d}{\mathrm{Re}^2} \qquad \text{(A.3)}$$

where $a = \dfrac{4P_{21}P_{12}p^2\beta^2 S}{(1-p)^2}$ , $b = \pi\beta(P_{22} - pP_{11})a_1$,

$d = a_3^2(\gamma_{1,1}^2 + 4\pi^2\beta^2 p^2)(\gamma_{1,2}^2 + 4\pi^2\beta^2)$ ,

$a_3 = \gamma_{1,2}^2 + \gamma_{1,1}^2 + 4\pi^2\beta^2(1 + p^2)$ , $c = a_3^2\beta^2\left(\pi^2(P_{22} - pP_{11})^2 - 4\dfrac{P_{12}P_{21}p^2 S ctg\pi(p + \frac{1}{p})}{(1-p)^2}\right)$.

2. In the limit $\mathrm{Re} \gg 1$, inequality (A.3) with $c > 0$ is reduced to the inequality

$$\mathrm{Re} > \mathrm{Re}_{th}(\beta) = \frac{\sqrt{c} - b \cdot \frac{S}{|S|}}{a} \qquad \text{(A.4)}$$

In (A.4), the function $\mathrm{Re}_{th}(\beta)$ takes minimal value (given in (6)) for

$$\beta = \beta_0 = \frac{1}{2\pi}\left[\frac{A(\gamma_{1,2}^2 + \gamma_{1,1}^2) + B(\gamma_{1,2}^2 - \gamma_{1,1}^2)}{A(1 + p^2) + B(1 - p^2)}\right]^{1/2} , \qquad \text{(A.5)}$$

where $A$ and $B$ are defined in the main text (see (6)) for $A^2 > 0$ .

3. On the neutral curve with $\lambda_1 = 0$ (i.e. when equality $\mathrm{Re} = \mathrm{Re}_{th}(\beta, p)$ in (A.4) holds), the phase velocity $V_\beta / V_{cp}$ has the form

$$V_\beta / V_{cp} = pP_{11} + P_{22} \pm \left(\frac{D_2^{1/2}}{2\sqrt{2\pi\beta}\,\mathrm{Re}}\right)_{\beta=\beta_{1,2}} \qquad \text{(A.6)}$$

where $\beta = \beta_{1,2}(\mathrm{Re}, p)$ corresponds to replacing of inequality in (A.4) by equality. For such a replacement in (A.4), we get a quadratic equation with respect to $\beta$. Its solution is

$$\beta = \beta_{1,2} = \frac{\mathrm{Re} \pm \sqrt{\mathrm{Re}^2 - \mathrm{Re}_{th}^2}}{2\pi^2 \delta_1} \qquad \text{(A.7)}$$

for $\mathrm{Re} \geq \mathrm{Re}_{th}$, where $\mathrm{Re}_{th}$ from (6), when $\beta_{1,2} = \beta_0$ for $\mathrm{Re} = \mathrm{Re}_{th}$, and

$$\delta_1 = \pi\left((p^2+1)\sqrt{1-4S\delta} - (1-p^2)\frac{Sb_1}{|Sb_1|}\right)\frac{|b_1|(1-p)^2}{|S|\,p^2 P_{12} P_{21}},$$

$$b_1 = P_{22} - pP_{11}, \delta = \frac{p^2 P_{21} P_{12}}{(1-p)^2 \pi^2 b_1^2}\, ctg\pi\left(p + \frac{1}{p}\right).$$

4.

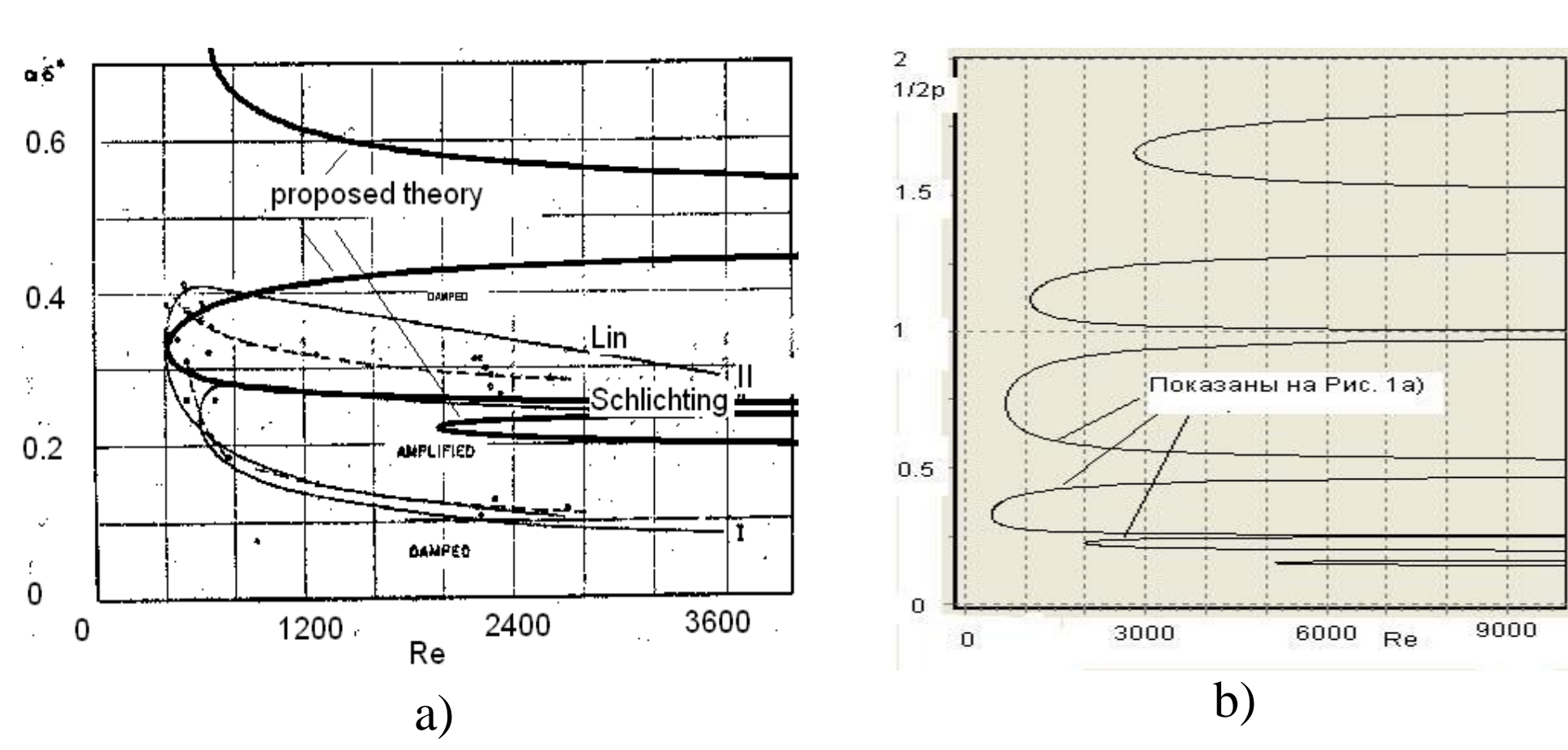

a) b)

Fig. 1. Family of the six curves of neutral stability (with $\lambda_1 = 0$), according to (6) and (A.5), b). On Fig..1a), scaled plots of three of them are given (they are noted also on Fig. 1b)). Meanwhile, the mean, second from below, instability region is bounded by the curve corresponding to the value $\beta_0$=0.463 (for $p$=1.527), and the lower one to $\beta_0$=1.099 (for $p$=2.239), according to (A.5). On Fig.1a), we give overlapping with a figure from [31] (see Fig.12 in [31]) under condition that formally, $1/2p = \alpha\delta^*$. In [31], $\alpha$ is the wave disturbance number, and $\delta^*$ is the boundary layer shift thickness when streamlining a thin plate. On Fig.1a, points and dashes correspond to the experiment [31], and the solid lines correspond to the Shlihting theory (the lower, denoted by I) and of Lin (the upper, denoted by II).

5. Table of values $\mathrm{Re}_{th}$ and $\beta_0$, obtained by (A.3) and (6) for $p$, corresponding to the local minima $\mathrm{Re}_{th}$ in the approximate formula (6)

| $p$ | $\beta_0$ (from (A.3)) | $\mathrm{Re}_{th}$ (from (A.3)) | $\beta_0$ (from (A.5)) | $\mathrm{Re}_{th}$ (from (6)) |
|---|---|---|---|---|
| 1,527 | 0,471 | 448,455 | 0,463 | 442,278 |
| 0,674 | 1,124 | 680,307 | 1,101 | 678,482 |
| 0,447 | 1,368 | 1095,455 | 1,358 | 1093,824 |
| 2,239 | 1,100 | 1983,171 | 1,099 | 1981,838 |
| 2,791 | 0,220 | 13095,398 | 0,219 | 13095,285 |
| 0,359 | 1,114 | 23816,499 | 1,114 | 23816,488 |

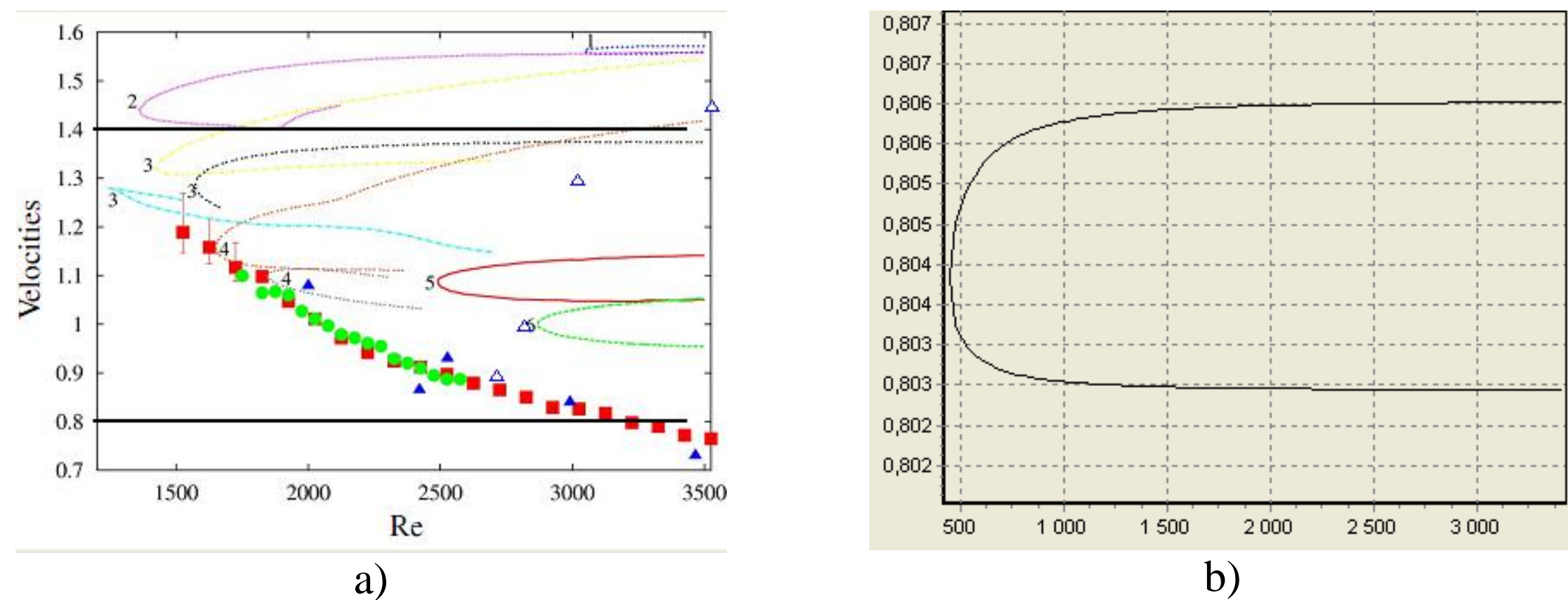

Fig. 2. a) Results of the experiment [28] (for the velocity of the rear edge of a turbulent spot, blue rectangles, and for the velocity of the fore edge, light triangles) and [30] (rectangles and circles, for the velocity of the rear edge). The phase velocity of the wave solutions [8] (numbers 1-5 denote the level of azimuthal symmetry of the corresponding travelling wave). The velocities are normalized by the stream in the pipe volumetric/average velocity. Also the result of calculations of the phase velocity according to (A.6) is given. The top "straight" line corresponds to sign plus in (A.6), and the bottom one to sign minus for $p \approx 1.53$, $\beta$ =0,471 (that corresponds to the absolute minimum $\mathrm{Re}_{th}$=448.5 according to (A.3). b) Zoomed representation of the bottom "straight" line from Fig.2,a according to (A.6) (the upper brunch on Fig.2,b corresponds to sign plus in (A.7), and the lower to sign minus).